\documentclass[12pt]{iopart}

\usepackage{graphicx}
\usepackage{color}
\usepackage{float}

\renewcommand{\vec}[1]{\mbox{\boldmath$\mathrm{#1}$}}

\begin{document}

\title{Quantum Otto heat engine based on a multiferroic chain working substance}

\author{M. Azimi$^{1,2}$, L.Chotorlishvili$^1$, S. K. Mishra$^{1,3}$, T. Vekua$^4$, W. H\"ubner$^5$, J. Berakdar$^1$}
\address{$^1$Institute of Physics, Martin-Luther University  Halle-Wittenberg, 06099  Halle, Germany}
\address{$^2$Max-Planck Institute for Microstructure Physics, Weinberg 2, 06120 Halle, Germany}
\address{$^3$Department of Physics, Indian Institute of Technology, Banaras Hindu University, Varanasi - 221005, India}
\address{$^4$Institute for Theoretical Physics, Leibniz University Hanover, 30167 Hanover, Germany}
\address{$^5$Department of Physics and Research Center OPTIMAS, University of Kaiserslautern, PO Box 3049, 67653 Kaiserslautern, Germany}
\ead{mazimi@mpi-halle.mpg.de}
%
%

\begin{abstract}
We study a quantum Otto engine operating on the basis of a helical spin-$\frac{1}{2}$ multiferroic chain with strongly coupled magnetic and ferroelectric order parameters. The presence of a finite spin chirality in the working substance enables steering of the cycle by an external electric field that couples to the electric polarization. We observe a direct connection between the chirality, the entanglement and the efficiency of the engine. An electric-field dependent threshold temperature is identified
 above which the pair correlations in the system, as quantified by the thermal entanglement, diminish.
  In contrast to the pair correlations, the collective many-body thermal entanglement is less sensitive to the electric field, and
   in the high temperature limit converges to a constant value. We also discuss the correlations between the threshold temperature of the pair entanglement, the spin chirality and the minimum of the fidelities in relation
    to the electric and magnetic fields. The efficiency of the quantum Otto cycle shows a saturation plateau with increasing electric field amplitude.
\end{abstract}


\maketitle
\section{Introduction}
With the advances in nanotechnology enabling a controlled miniaturization and a functionalization of  nanostructured materials, questions
related to the thermodynamical properties are gaining an increased attention.  Several theoretical proposals were put forward
for nanoscale  Brownian motors \cite{Humphrey}, refrigerators \cite{Skrzypczyk} and quantum heat engines \cite{Henrich,Quan1,Quan2,Abah,Linden,Wang1,Jarzynski,Deffner,Campo,Wang2,Wang3}. On the other hand,
 for finite systems, the application of the laws of thermodynamics is the subject of an ongoing debate \cite{Campisi}.
One of the fundamental questions concerns the size limit to which the working substance might be scaled down.
Recent studies point out that the quantum nature of a size-quantized working substance, e.g.
 a quantum heat engine, may lead to a close connection between the efficiency of the cycle and quantum correlations \cite{Dillenschneider}, which can be quantified in terms of the entanglement \cite{Wootters,Amico1,Mintert}, a behavior that is atypical
for classical engines.
 According to the fundamental laws of thermodynamics, the efficiency of a classical engine is independent of its detail
  and is solely determined by the character of the cycle itself and the temperatures of the heat baths.
  The quantum nature of the working substance, however, has key consequences for the engine output power as well.
  Recently it was shown that purely quantum phenomena, such as noise-induced coherence, yields greater engine output power \cite{Scully1,Scully2}. \\
  In general, physical phenomena at the crossover of quantum mechanics and thermodynamics are the subjects of the emergent field
  of quantum thermodynamics where, among other topics, questions are addressed as to what extent standard
classical thermodynamic cycles, such as Carnot or Otto cycles, can be reformulated for quantum systems \cite{Quan1}.
A key issue thereby is the difference between thermodynamic and quantum adiabatic processes.
 For example, a thermodynamical  adiabatic process does not necessarily mean that
the occupation probabilities are invariant during an adiabatic transition.
As usual, thermodynamical adiabatic processes are identified
in terms of the conservation of the entropy and the isolation of the system from the heat exchange with the thermal bath.
 An essential requirement for the  quantum adiabatic
 process is that the population distributions  remain unchanged. Thus, quantum adiabaticity is a stricter requirement
 than the thermodynamic one.  The adiabatic quantum process is also adiabatic in the thermodynamic sense, the opposite is however not true in general. Therefore,
  quantum adiabaticity entails a relatively low power output from a slowly operating quantum engine, unless the
  energy spectrum of the working substance has
 nodal crossing points, however, this is not a generic feature of realistic physical systems.
  Landau Zener transitions are avoided
during an adiabatic segment of the cycle by  slowly varying the control parameters \cite{Shevchenko}. As  mentioned
 above, the subtlety of quantum engines is related to the internal connection between essentially quantum phenomena such as entanglement and the thermodynamic
characteristics of the cycle. In this regard, the choice of the working substance for the operating quantum engine is an important issue \cite{Chotorlishvili}.\\
Recently, there has been a great interest in  composite multiferroic (MF) materials that  possess  coupled ferromagnetic (FM) and ferroelectric
(FE) properties \cite{Wang4,Eerenstein,Garcia,Bibes,Cheong,Dawber,Valencia,Duan,Meyerheim,Horley,Sahoo,6,Fechner,Hearmon,Menzel} (for a review we refer to \cite{advaphys}). These
materials allow for a multitude of novel applications based on the control of magnetism (ferroelectricity) with electric (magnetic) fields.
 They offer new opportunities for the design and control of new circuits for quantum information processing.
 For an interesting class of magnetoelectrics, the ME coupling is rooted in a chiral magnetic ordering that is coupled
 to an electric polarization $\vec{P}$ such that\cite{Mostovoy}
\begin{eqnarray}\label{eq.1}
\displaystyle\vec{P}\sim\vec{r}_{i,i+1}\times\big(\vec{\sigma}_{i}\times\vec{\sigma}_{i+1}\big).
\end{eqnarray}
Above  $\vec{r}_{i,i+1}$, is the relative spatial vector between the effective spins $\vec{\sigma}_i$ and $\vec{\sigma}_{i+1}$ localized at neighboring sites.
Though Eq.(\ref{eq.1}) was derived initially phenomenologically, a fully microscopic theory based on the electronic states was developed shortly thereafter \cite{Katsura}.
 The emergence of an electric polarization, when coupling the spatial degrees of freedom to the
 spin chirality, renders possible an efficient  manipulation and control of the spin order parameter via an applied external electric field.

In the present project, we will study a model
for a quantum Otto engine operating on the basis of a one-dimensional (1D) finite size MF chiral spin chain that acts as a working substance.
The possibility of controlling the efficiency of the quantum heat engine via an external electric field motivates our choice of the working substance.
For small working substance consisting of four spins with periodic boundary conditions, we provide an analytical solution to the problem.
For a larger size of the working substance an exact numerical diagonalization reveals the connection between the thermal entanglement and the cycle efficiency.

\begin{center}
\begin{figure}[H]
\centering
\includegraphics*[width=0.65\textwidth]{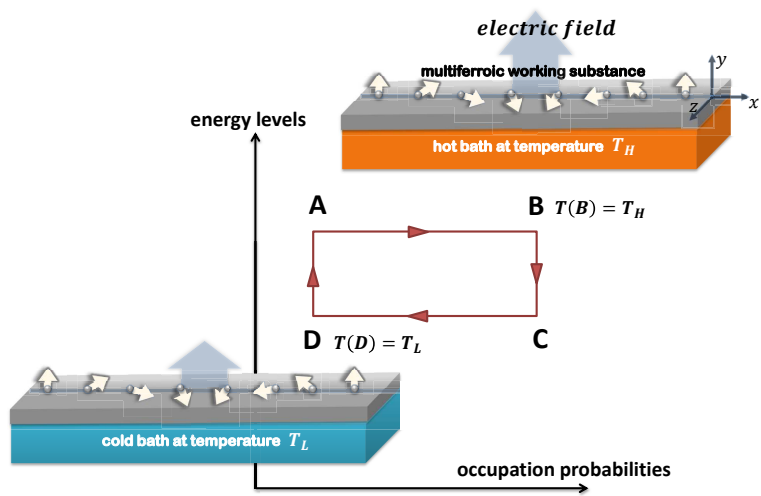}
\caption{(Color online)
A schematic of the considered quantum Otto cycles based on a chiral multiferroic chain.
The  cycle has four strokes: Step $A\rightarrow B$ and $C\rightarrow D$ are two isochoric processes.
During the step $A\rightarrow B$ the system is attached to a hot bath with a temperature $T_{H}$.   Step
$C\rightarrow D$ is inverse to the step $A\rightarrow B$. After releasing (absorbing) energy to the cold (hot) heat bath with temperature $T_{L}$
($T_{H}$) the system reaches a thermodynamic equilibrium state associated with the level populations $P_{n}^{D}\big(E_{n}(\wp_{1}),T_{L}\big)$
($P_{n}^{B}\big(E_{n}(\wp_{}),T_{H}\big)$.
 Steps $B\rightarrow C$ and $D\rightarrow A$ are two thermodynamic adiabatic processes. During the process $B\rightarrow C$ the amplitude of the electric field is changed so that  $\Delta E_{n}=E_{n}(\wp)-E_{n}(\wp_{1})$ and the working substance performs a positive work.}
\label{system}
\end{figure}
\end{center}

Our theoretical model is experimentally feasible. Recently discovered materials such as the quantum $S=1/2$
spin chain magnets $LiCu_{2}O_{2},~~CoCr_{2}O_{4},~~LiVCuO_{4}$, possess simultaneously ferroelectric and ferromagnetic properties. $1D$ patterns of the chain magnets can be manufactured using $CuO_{2}$ powders and a $Pt$ stove \cite{Park}. Then, a spin chain magnet  doped on the $Ir (001)$
or nonmagnetic $Zn^{2+}$ substrate could serve as a working substance, while a $Pt$ stove could be implemented as a thermal bath for controlling the temperature of the working substance.  The cycle we are going to study consists of two thermodynamic adiabatic and two isochoric strokes. During the two isochoric strokes, the multiferroic spin chain interacts with the heat baths. During the two thermodynamic adiabatic strokes the amplitude of the electric field is changed (cf. Fig.\ref{system} for
a schematic illustration). Since the energy levels of the system depend on the electric field, a change in the electric field modifies the energy levels. In this way work is done by the engine during  the thermodynamic adiabatic strokes. In what follows, we will study the connection between the cycle efficiency, the entanglement and the electric field.

The paper is organized as follows. In section II we introduce our model. In section III we present analytical results obtained for the multiferroic chain working substance consisting of four spins. In particular we study: the dependence of the pair and the nonlocal many-body entanglements on the temperature and the amplitude of the electric field and the temperature dependence of the quantum chirality and the electric and magnetic susceptibilities. Each of these quantities has  a particular meaning: Quantum chirality is a measure for the spin frustration and allows to drive the cycle and to control the cycle efficiency by an external electric field. The electric and magnetic susceptibilities can be used to detect the thermal phase transitions in the working substance, and the local and many-body entanglements are useful to observe the connection between the cycle efficiency and the quantum correlations. In section IV we discuss details of the thermodynamic cycle and in section V we evaluate the scaling of the cycle efficiency with the size of the working substance. In section VI we study the quantum Otto cycle in the semiclassical limit and wrap up in section VII.

\section{Model}

We envisage the application of a MF helical chain in one spatial dimension as an electric-field controlled heat engine.
For definiteness we take the $x$ axis as the chain direction.
 An effective model that captures the physics of the MF chain \cite{Katsura,Azimi} is based on the Hamiltonian
\begin{eqnarray}\label{eq.2}
\hat{H}=-J_1\displaystyle\sum_{i=1}^{N}\vec{\sigma}_i.\vec{\sigma}_{i+1}
-J_2\displaystyle\sum_{i=1}^{N}\vec{\sigma}_i.\vec{\sigma}_{i+2}
-\gamma_{e}\hbar B\displaystyle\sum_{i=1}^{N}\sigma_{i}^{z}-\vec{\wp}\vec{P}.
\end{eqnarray}
We assumed here that the chain is subjected to an electric field ($\vec{\wp}=(0,\wp,0)$  applied along the $y$ axis
 and to a magnetic $B$ along the $z$ axis.
The  exchange interaction constant between the nearest neighbor spins is chosen ferromagnetic $J_1>0$ while the next-nearest interaction is antiferromagnetic $J_2<0$.
Pauli matrices are used in standard notations $\vec{\sigma}_i$, and $\gamma_{e}$ is the gyromagnetic ratio for electron spin.
 $\hbar$ is  Planck's constant.
With the help of Eq.(\ref{eq.1}) the coupling of the electric field to the MF chain can be written as $\vec{\wp}\vec{P}=\wp g_{ME} \sum_{i}(\vec{\sigma}_{i}\times
\vec{\sigma}_{i+1})_{z}$, where $g_{ME}$ is the magnetoelectric coupling strength. The quantity $\kappa_i=(\vec{\sigma}_{i}\times
\vec{\sigma}_{i+1})_{z}$ is known as the $z$ component of the vector chirality (VC) (that we will simply call chirality).
Electric field coupling resembles the Dzyaloshinskii-Morija (DM) anisotropy, with the constant $d=\wp g_{ME}$.
The effective model Hamiltonian (\ref{eq.2}) is relevant for $1D$ spin frustrated MF copper oxides $LiCu_{2}O_{2},~~CoCr_{2}O_{4},~~LiVCuO_{4}$ as discussed in the literature \cite{Yamasaki,Rusydi,Yasui,Loidl}. For $LiCu_{2}O_{2}$ the values of the model parameters are $J_{1}\approx 81K,~~J_{2}\approx 44K$ see \cite{Yasui}.

The problem of an electric-field control of the magnetic chirality of a ferroaxial MF system was addressed in a recent paper \cite{Hearmon}. Information transfer by the vector spin chirality in magnetic chains was discussed in \cite{Menzel}. The effect of the electric field, or the DM anisotropy variation  on the quantum information processing as well as on many-body quantum ground states and quantum-phase transitions
 of MF helical chain we addressed recently in \cite{Azimi}.

In what follows, we suppose that $J_{1}=-J_{2}=J$ and go over to dimensionless units such that
$B\rightarrow\gamma_{e}\hbar B/J,~~\wp\rightarrow g_{ME}|\overrightarrow{\wp}|/J$, i.e.,
 we measure  the Zeeman energy and  the interaction energy with electric field in units of the exchange constant.
 As was mentioned in the introduction, the purpose of the present project is to investigate a possible control of  the cycle efficiency and thermal entanglement via an external electric driving field. For clarity we combine analytical and full numerical approaches and start with a solvable model consisting of a four spins with periodic boundary conditions as a working substance.

\section{Four spins case: Analytical treatment}

In the case of four spins, the Hamiltonian (\ref{eq.2}) can be diagonalized analytically.
All technical details are provided as supplementary materials to this paper.
 Here we present the  main results.
To study the thermal entanglement and the cycle efficiency we construct the density matrix $\hat{\rho}$ corresponding to an equilibrium Gibbs distribution,
\begin{eqnarray}\label{eq.3}
\hat{\rho}=Z^{-1}\sum_{n=1}^{16}\exp[-\beta E_n]\vert\psi_n\rangle\langle\psi_n\vert,~~~~Z=\sum_{n=1}^{16}\exp[-\beta E_n],
\end{eqnarray}
where $\vert\psi_n\rangle$ and $E_n$ are the eigenfunctions and eigenvalues of (\ref{eq.2}) given explicitly  in the supplementary materials (See (A1), (A2)). Using the
density matrix (\ref{eq.3}) we calculate the mean value of the $z$ component of the VC. The only nonzero component for the considered configuration of the system and the chosen direction of the electric field is  $\langle\displaystyle\sum_{i=1}^{4}[\hat{e}_x\times(\vec{\sigma}_{i}\times
\vec{\sigma}_{i+1})]_{y}\rangle=\tr(\hat{\rho}\sum_{i=1}^{4}[\hat{e}_x\times(\vec{\sigma}_{i}\times
\vec{\sigma}_{i+1})]_{y})$,
where $\tr(\cdots)=\sum_{n=1}^{16} \langle \psi_n\vert \cdots  \vert\psi_n\rangle $.

Following standard definitions \cite{Amico1} we calculate the pair concurrence between  two  arbitrary spins of the working substance $C_{nm}=max(0,\sqrt{R_{nm}^{(1)}}-\sqrt{R_{nm}^{(2)}}-\sqrt{R_{nm}^{(3)}}-\sqrt{R_{nm}^{(4)}})$. Here $R^{(\alpha)}_{nm}$, $\alpha=1,2,3,4$ are the eigenvalues of the matrix
$R_{nm}=\rho_{nm}^{R}(\sigma_{1}^{y}\bigotimes\sigma_{2}^{y})(\rho_{nm}^{R})^{*}(\sigma_{1}^{y}\bigotimes\sigma_{2}^{y})$
and $\rho_{nm}^{R}$ is the reduced density matrix of the system of two spins obtained from the density matrix of the system $\hat{\rho}$ (\ref{eq.3}) after tracing out two remaining spins $\rho_{nm}^{R}=\tr_{sp}(\hat{\rho})$, where $s,p \neq m,n$.  After some rather straightforward calculations we obtain

\begin{eqnarray}\label{eq.4}
R_{12}=\frac{1}{Z^2}\left(
                \begin{array}{cccc}
                  a_1d_1 & 0 & 0 & 0 \\
                  0 & b_{1}^{2}+{|c_1|}^2 & 2b_1c_1 & 0 \\
                  0 & 2b_1c_{1}^{*} & b_{1}^{2}+{|c_1|}^2 & 0 \\
                  0 & 0 & 0 & a_1d_1 \\
                \end{array}
              \right).
\end{eqnarray}

Similarly we calculate for the other components

\begin{eqnarray}\label{eq.5}
R_{13}=\frac{1}{Z^2}\left(
                \begin{array}{cccc}
                  a_2b_2 & 0 & 0 & 0 \\
                  0 & c_{2}^{2}+{|d_2|}^{2} & 2c_2d_2 & 0 \\
                  0 & 2c_2d_{2}^{*} & c_{2}^{2}+{|d_2|}^2 & 0 \\
                  0 & 0 & 0 & a_2b_2 \\
                \end{array}
              \right),
\end{eqnarray}

\begin{eqnarray}\label{eq.6}
R_{14}=\frac{1}{Z^2}\left(
                \begin{array}{cccc}
                  a_1d_1 & 0 & 0 & 0 \\
                  0 & b_{1}^{2}+|c_1|^2 & 2b_1c_{1}^{*} & 0 \\
                  0 & 2b_1c_1 & b_{1}^{2}+{|c_1|}^2 & 0 \\
                  0 & 0 & 0 & a_1d_1 \\
                \end{array}
              \right).
\end{eqnarray}

In view of Eqs. (\ref{eq.4})-(\ref{eq.6}) for the different pair concurrences we infer

\begin{eqnarray}\label{eq.7}
&&C_{12}=C_{14}=\frac{2}{Z}max\{|c_1|-\sqrt{a_1d_1},0\},\nonumber
\\
&&C_{13}=\frac{2}{Z}max\{|d_2|-\sqrt{a_2b_2},0\}.
\end{eqnarray}

Explicit expressions of the parameters $a_{1,2},b_{1,2},c_{1,2},d_{1,2}$ are quite involved and therefore are presented in the supplementary materials (see Eqs. (A3), (A4)).
The pair concurrences $C_{nm}$ depend on the chosen spins $n$ and $m$ (more precisely, due to the translational symmetry of periodic chain, on the distance between the two spins) and as one can see from Eq.(\ref{eq.7}), $C_{nm}$ are quite different from each other. Therefore,
 the more informative and universal quantity seems to be the two tangle \cite{Roscilde} $\tau_2$ which contains information on the total pair correlations in the spin chain $\tau_2=2C_{12}^{2}+C_{13}^{2}$. Taking into account Eq.(\ref{eq.7}) for the two tangle we deduce

\begin{eqnarray}\label{eq.8}
&&|c_1|>\sqrt{a_1d_1},|d_2|>\sqrt{a_2b_2}: \,\,\, \tau_{2}=\frac{8{(|c_1|-\sqrt{a_1d_1})}^2+4{(|d_2|-\sqrt{a_2b_2})}^2}{{Z}^2}, \nonumber
\\
&&|c_1|>\sqrt{a_1d_1},\,\,|d_2|<\sqrt{a_2b_2}:\,\,\,   \tau_{2}=\frac{8{(|c_1|-\sqrt{a_1d_1})}^2}{{Z}^2},
\\
&& |c_1|<\sqrt{a_1d_1},|d_2|>\sqrt{a_2b_2}:\,\,\, \tau_{2}=\frac{4{(|d_2|-\sqrt{a_2b_2})}^2}{{Z}^2},\nonumber
\\
&&|c_1|<\sqrt{a_1d_1},|d_2|<\sqrt{a_2b_2}: \,\,\, \tau_{2}=0.\nonumber
\end{eqnarray}

The degree of the pair correlations depends thus on several inequalities between the parameters $a_{1,2},b_{1,2},c_{1},d_{1,2}$ which are functions of the temperature $T$, and the amplitudes of the driving electric and magnetic fields  $\wp,B$. The explicit expressions of the parameters entering in Eqs.(8) can be found in the supplementary materials as Eqs. (A3) and (A4). In effect, depending on the values of the three parameters $\wp,B,T$ the system can be entangled or disentangled. The threshold temperature $T_{c}(\wp,B)$  for the given amplitudes of the driving fields defines the regimes  of  entangled and disentangled states.
 Our principle interest now is to see how the threshold temperature of the system scales with the electric field $T_{c}(\wp)$.

Another interesting object is the one tangle \cite{Roscilde} $\tau_1=4det\rho_1$ as it quantifies the nonlocal many-body correlations in the spin chain. Here $\hat{\rho_1}=\tr_{2,3,4}(\hat{\rho})$ is the reduced density matrix of the first spin after tracing out the states of all other spins. Explicit expressions for the one tangle can be found in the analytical form $\tau_1=\frac{4}{Z^2}Q$, where the explicit form of $Q$ is presented in the supplementary materials (see (A5)). One tangle and the collective nonlocal entanglement is related to the complex spiral spin structure of the frustrated MF chain and therefore is generated by the external electric field.
An indication of the existence of the spiral spin structure in the system is the nonzero chirality.
 Therefore, the values of the $z$ component of the vector chirality $\tr(\hat{\rho}\sum_{i=1}^{4}[\hat{e}_x\times(\vec{\sigma}_{i}\times
\vec{\sigma}_{i+1})]_{y})$ and the values of the one tangle $\tau_1=\frac{4}{Z^2}Q$  are in correlation with each other. Both of them depend on the amplitude of the electric field. Based on the definition of the parameter $Q$ (supplementary materials Eq.(A5)) it is easy to see that in the high temperature limit $T\rightarrow \infty $ for one tangle we obtain $\tau_1=1$. Considering  the eigenvalues and eigenfunctions of the Hamiltonian (\ref{eq.2}) (see supplementary materials Eqs. (A1) and (A2)) for the chirality of the MF chain we obtain:

\begin{eqnarray}\label{eq.9}
&&\langle\displaystyle\sum_{i=1}^{4}[\hat{e}_x\times(\vec{\sigma}_{i}\times
\vec{\sigma}_{i+1})] _{y}\rangle=\frac{4}{Z}(e^{-\beta E_2}-e^{-\beta E_3}+8\alpha^2\mu e^{-\beta E_6}\nonumber
\\
&&~~~~~~~~~~~~~~~~~~~~~~~~~~~~~~~~~~~~~~~~~~~+8\gamma^2\lambda e^{-\beta E_7}-e^{-\beta E_{12}}+e^{-\beta E_{13}}).
\end{eqnarray}

\begin{center}
\begin{figure}[H]
\centering
\includegraphics*[width=0.55\textwidth]{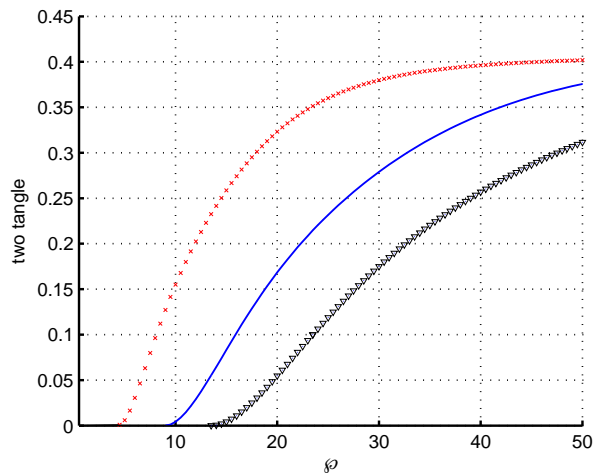}
\caption{(Color online) Dependence of the two tangle $\tau_{2}(\wp)$ on the electric field for $B=1$, and for three different Temperatures: red cross line $T=10$, blue solid line $T=20$, black triangular line $T=30$.}
\label{Fig.1}
\end{figure}
\end{center}

 The magnetic field dependence of both types of the entanglement, namely the short-range pair correlations as quantified by $\tau_{2}$ and the many-body collective entanglement described by $\tau_{1}$, is transparent (this is due to the fact that in our model
  the magnetic field couples to the magnetization which is a conserved quantity in our model).
   With  increasing  $B$  the entanglements decrease. The dependence on the amplitude of the electric field is less obvious
 and deserves a detailed consideration. First we focus on the  thermal pair entanglement $\tau_{2}$. As we see in  Fig. \ref{Fig.1},
$\tau_2$ is finite only for a very large amplitude of the electric field.

\begin{center}
\begin{figure}[H]
\centering
\includegraphics*[width=0.55\textwidth]{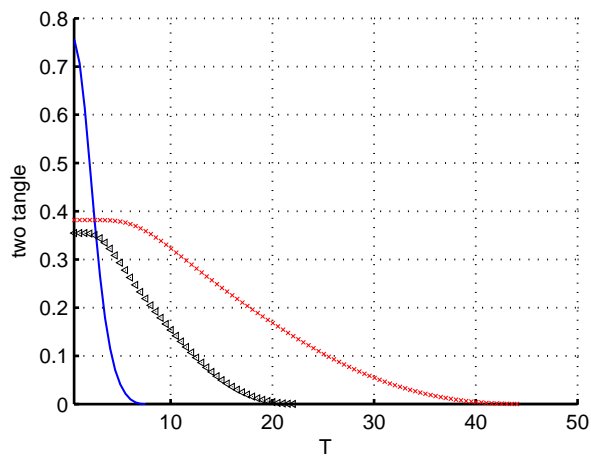}
\caption{(Color online) Dependence of the two tangle $\tau_{2}(T)$ on the temperature for $B=1,$ and for  fixed values of the electric field:
blue solid line $\wp=1$, black triangular line $\wp=10$, red cross line $\wp=20$
The threshold temperatures are $T_{c}=7.37,~~T_{c}=22.31,~~T_{c}=44.45$, respectively.}
\label{Fig.2}
\end{figure}
\end{center}

The pair thermal concurrence $\tau_{2}$ is practically zero until the electric field amplitude becomes quite substantial as can be seen in Fig. \ref{Fig.1} (recall we are operating in scaled units).
We also determine the threshold temperature below which $\tau_{2}$ is finite and above which $\tau_{2}=0$, see Fig. \ref{Fig.2}.

\begin{center}
\begin{figure}[H]
\centering
\includegraphics*[width=0.55\textwidth]{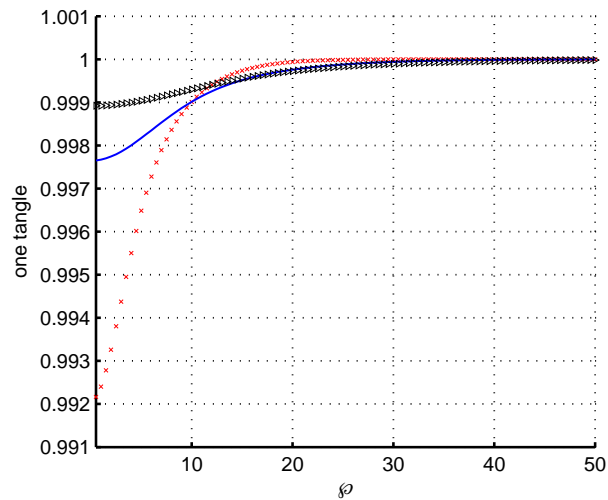}
\caption{(Color online) Dependence of the one tangle $\tau_{1}(\wp)$ on the electric field for $B=1$, and for three different temperatures: red cross line $T=10$, blue solid line $T=20$, black triangular line $T=30$.}
\label{Fig.3}
\end{figure}
\end{center}

In contrast  to the pair correlations and the entanglement, the collective entanglement $\tau_{1}$ is different from zero for an arbitrary
electric field. (see Fig. \ref{Fig.3}).

\begin{center}
\begin{figure}[H]
\centering
\includegraphics*[width=0.60\textwidth]{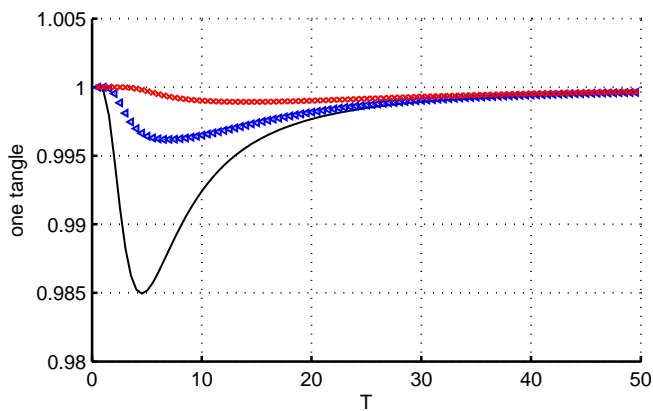}
\caption{(Color online) Dependence of the one tangle on the temperature for  $B=1,$ and fixed values of the electric field: black solid line $\wp=1$, blue triangular line $\wp=5$, Red cross line $\wp=10$.}
\label{Fig.4}
\end{figure}
\end{center}

Another remarkable difference is that  the collective entanglement $\tau_{1}$ is very robust and
is practically not effected by the temperature (see Fig. \ref{Fig.4}).

\begin{center}
\begin{figure}[H]
\centering
\includegraphics*[width=0.55\textwidth]{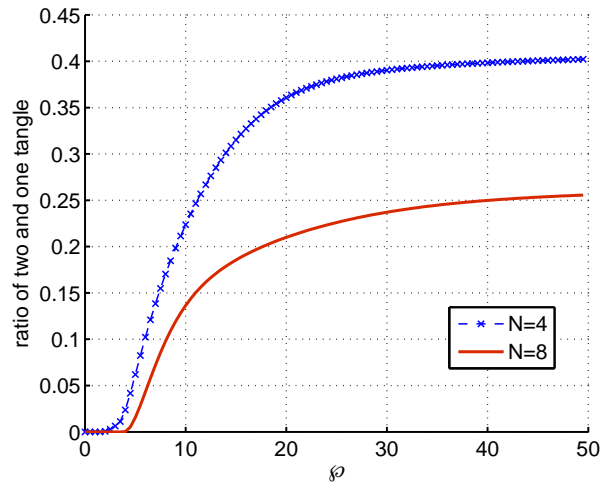}
\caption{(Color online) Ratio between two and one tangle $\tau_{2}/\tau_{1}$ as a function of external electric field for
the following values of the parameters $B=1,~T=7.37$. Result for $N=4$ spins is plotted using obtained analytical
solutions. Result for $N=8$ spins is plotted using numerical solutions.}
\label{Fig.5}
\end{figure}
\end{center}

Therefore, the amount of the thermal entanglement stored in
the nonlocal correlations  $\tau_{1}$ is always larger than the
thermal entanglement of the pair correlations $\tau_{1}>\tau_{2}$, as shown in Fig. \ref{Fig.5}.

\begin{center}
\begin{figure}[H]
\centering
\includegraphics*[width=0.55\textwidth]{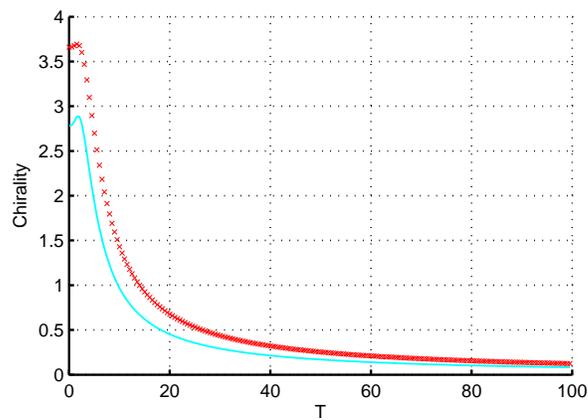}
\caption{(Color online) Chirality as a function of the temperature, for  $B=1$, and for the fixed values of the electric field: cyan solid line $\wp=1$, red cross line $\wp=1.5$.}
\label{Fig.6}
\end{figure}
\end{center}

With  increasing  the size of the working
substance the ratio between $\tau_{2}/\tau_{1}$ becomes smaller. This means that the many-body  entanglement $\tau_{1}$ is increasing
with the size of the system $N$ faster than the total two pair correlations $\tau_{2}$.
The situation with respect to the thermal chirality is different. In particular, we observe that with
the increase of the temperature the thermal chirality undergoes a strong change and above the threshold temperature $T_{c}$ of the two tangle $\tau_{2}$, the thermal chirality is almost zero, as depicted in
Fig. \ref{Fig.6}.

\begin{center}
\begin{figure}[H]
\centering
\includegraphics*[width=0.55\textwidth]{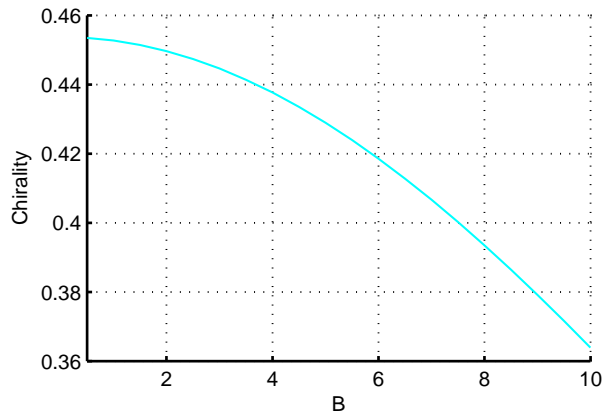}
\caption{(Color online) Chirality as a function of the magnetic field for the following values of the parameters $\wp=1,~T=20$.}
\label{Fig.7}
\end{figure}
\end{center}

Thus, we see that the thermal chirality is  correlated with the pair correlations in the system. As for the temperature,
the magnetic field also has a negative influence on the chirality.
The dependence of the thermal chirality on $B$ is plotted in Fig. \ref{Fig.7}, where one can see that with increasing  $B$,  the
thermal chirality decreases.

Another quantity of interest that quantifies the sensitivity to perturbations of quantum systems near a critical region is the fidelity \cite{Zanardi1}. A zero temperature fidelity is a measure for the overlap between two ground states corresponding to slightly different values of the controlling parameters. A dip  in fidelity reflects changes in the structure of the ground state at a quantum critical point \cite{Zanardi1}. The finite temperature thermal state extension of the quantum fidelity was considered in \cite{Zanardi2}. The fidelity of a mixed state at a finite temperature characterizes a second-order thermal phase transition \cite{Quan2} and is defined in the following way

\begin{eqnarray}\label{eq.10}
F_{\zeta}(\beta,\zeta_{0},\zeta_{1})=
\tr\sqrt{\sqrt{\hat{\rho_{0}}}\hat{\rho_{1}}\sqrt{\hat{\rho_{0}}}},
\end{eqnarray}
where $\hat{\rho_{0}}(\beta,\zeta_{0}),\hat{\rho_{1}}(\beta,\zeta_{1})$  are the density matrixes of the system corresponding to  slightly different control parameters $\zeta_1=\zeta_{0}+\delta\zeta$ and  $\beta=\frac{1}{k_{\beta}T}$. The expression for the  fidelity related to the electric and magnetic fields can be simplified to the following form

\begin{eqnarray}\label{eq.11}
F_{\zeta}(\beta,\zeta,\zeta+\delta \zeta)=\exp\bigg[-\frac{\beta{(\delta \zeta)}^{2}}{8}\chi(\zeta)\bigg],
\end{eqnarray}
where $\chi(\zeta)=-\frac{\partial^{2}F}{\partial \zeta^2}$ is the susceptibility to the corresponding external field  $\zeta= \wp, B$ at constant temperature. Analytical expressions of the susceptibilities are presented in the supplementary materials (Eq. (A6)-(A7)).

\begin{center}
\begin{figure}[H]
\centering
\includegraphics*[width=0.55\textwidth]{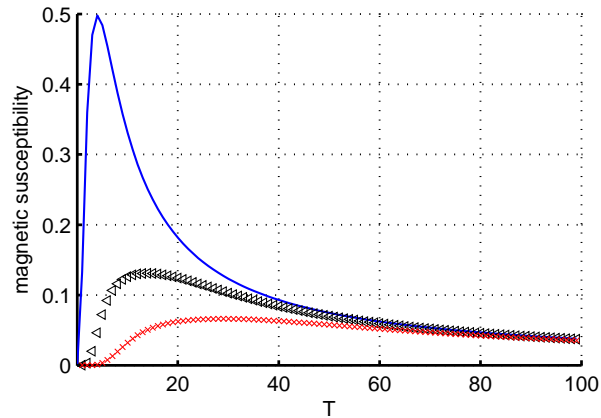}
\caption{(Color online) Magnetic susceptibility as a function of  the temperature for  $B=1$ and for the fixed values of the electric field: blue solid line $\wp=1$, black triangular line $\wp=10$, red cross line $\wp=20$. We see that with the increasing of the electric field amplitude the
 maximum of the magnetic susceptibility is shifted towards higher temperatures.}
\label{Fig.8}
\end{figure}
\end{center}

\begin{center}
\begin{figure}[H]
\centering
\includegraphics*[width=0.55\textwidth]{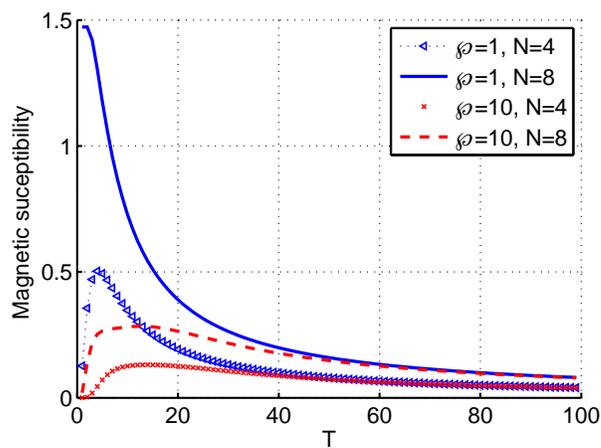}
\caption{(Color online) The magnetic susceptibility is plotted as a function of  the temperature for $N=4$ and $N=8$ cases. We choose $B=1$. The peak in the susceptibility shifts towards  lower temperatures
as we increase the system size.
 A similar peak in the magnetic susceptibility at finite temperatures was observed for $B=\wp=0$ and $N=24$ in \cite{Honecker}  and was interpreted as the result of a  competition between antiferromagnetic and ferromagnetic correlations in the system.}
\label{Fig.8N}
\end{figure}
\end{center}

\begin{center}
\begin{figure}[H]
\centering
\includegraphics*[width=0.55\textwidth]{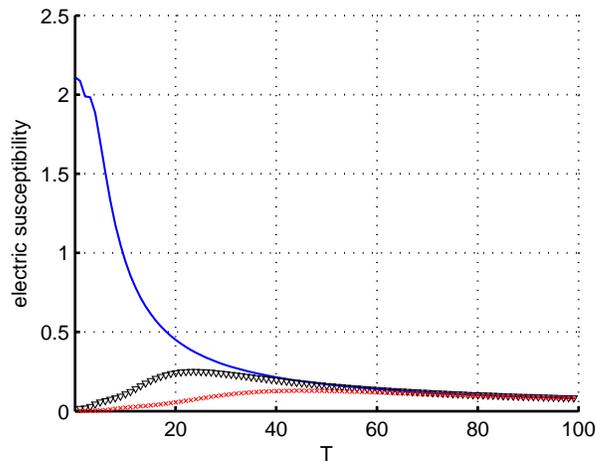}
\caption{(Color online) The electric susceptibility as a function of  the temperature for the following values of the parameters $B=1$ and for  fixed values of the electric field: blue solid line $\wp=1$, black triangular line $\wp=10$, red cross line $\wp=20$. Comparing this result to  Fig.\ref{Fig.3}  we see that with the increase  of the threshold temperature of pair correlations $\tau_{2}$  the maximum of the electric susceptibility drifts towards higher temperatures. $T_{c}\approx24,~~T_{c}\approx45$.}
\label{Fig.9}
\end{figure}
\end{center}

\begin{center}
\begin{figure}[H]
\centering
\includegraphics*[width=0.55\textwidth]{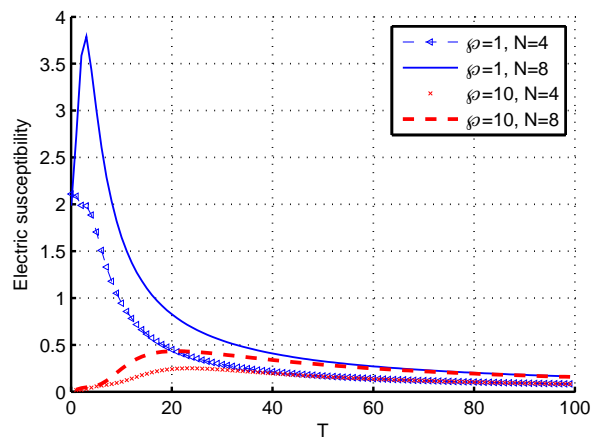}
\caption{(Color online) The electric susceptibility as a function of the temperature for $N=4$ and $N=8$ cases. we set  $B=1$. Note, the electric susceptibility increases as the
system size increases.}
\label{Fig.9N}
\end{figure}
\end{center}

Finally,  the dependence of the electric $\chi\big(\wp\big)$ and the magnetic $\chi\big(B\big)$ susceptibilities on the temperature are depicted in Fig. \ref{Fig.8}, Fig. \ref{Fig.8N}, Fig. \ref{Fig.9} and Fig. \ref{Fig.9N}.

According to the definition Eq.(\ref{eq.11}), the maxima of the susceptibilities correspond to the minima in the fidelities that are related to the electric and magnetic fields. Comparing Fig. \ref{Fig.8} - Fig. \ref{Fig.9N} with Fig. \ref{Fig.2}, we see a direct correlation between the threshold temperature of the pair entanglement $\tau_{2}$ and the minima of the fidelities related to the electric and magnetic fields. The maxima of the electric and magnetic susceptibilities are related to threshold temperatures of the pair correlations.
 For larger threshold temperatures of the pair correlations, the maxima of the electric and magnetic susceptibilities are shifted towards higher temperatures. Interestingly for the electric susceptibility, the correlation between the threshold temperature of the pair entanglement $\tau_{2}$ and the minimum of the fidelity is not only qualitative but also quantitative as well. As we see for large enough electric fields the maximum of the electric susceptibility is observed almost on the threshold temperatures $~~T_{c}\approx24,~~T_{c}\approx45$ of the pair entanglement $\tau_{2}$.

In Figs. \ref{Fig.8N} and \ref{Fig.9N} we present the system size dependence of the electric and the magnetic susceptibilities. The heights of the peaks of the electric and the magnetic susceptibilities increase with the system size.
 One can as well observe that for $N=8$ the location of the peak of the magnetic (electric) susceptibility shifts towards lower (higher) temperatures.

\section{Efficiency of the multiferroic heat Otto engine}
\label{efficiency_MF}
In analogy to the classical Otto cycle, the quantum Otto cycle also consists of  two quantum isochoric and two  adiabatic
processes \cite{Quan1}, as sketched in Fig.\ref{system}. The quantum isochoric process corresponds to a heat exchange between the working substance and  cold and hot heat baths.
During the quantum isochoric process only level populations $P_{n}$ are reshuffled, while during the adiabatic process the working substance produces work and in this case the energy levels are changed. Therefore, the work produced by the engine depends on the  amplitude of electric field which causes changes in energy levels.
An adiabatic process can be thermodynamic adiabatic or quantum adiabatic. A process is thermodynamic adiabatic if the working substance is thermally isolated from the heat exchange with the heat bath. However, this does not exclude  inter-level transitions of a purely quantum nature,
while in the case of a quantum adiabatic process the  level populations are fixed.

As was mentioned in introduction, in our case the working substance is a MF spin frustrated
chain with a discrete energy spectrum of 16 levels. The first law of thermodynamics for a system with discrete energy spectrum reads
\begin{eqnarray}\label{eq.12}
dU\big(E_{n},T\big)=\displaystyle\sum_{n=1}^{16}\bigg(E_{n}dP_{n}+P_{n}dE_{n}+E_{n}\bigg(\frac{\partial P_{n}}{\partial E_{n}}\bigg)_{T=const}dE_{n}\bigg).
\end{eqnarray}
Here $dU$ is the change of the system energy $U\big(E_{n},T)=\tr\big(\hat{\rho}\hat{H}\big)=\displaystyle\sum_{n=1}^{16}E_{n}P_{n}\big(E_{n},T\big)$.
The first term on the right hand side of Eq. (\ref{eq.12}), $\delta Q=E_{n}dP_{n}$ can be viewed  as the heat exchange and is related to the change of the level populations  $P_{n}\big(E_{n},T)$ occurring due to a change of the temperature for $E_{n}=const$, while the second and the third terms correspond to the produced work. If the adiabatic strokes of the cycle are quantum adiabatic then $\bigg(\frac{\partial P_{n}}{\partial E_{n}}\bigg)_{T=const}=0$ and Eq.(\ref{eq.12}) reduces to the form given in \cite{Quan1}. The work produced during the quantum adiabatic process reads $\delta W=P_{n}dE_{n}$. The working substance produces work due to the change of the amplitude of electric field $\wp$.
This leads to a modification in the energy levels with $\Delta E_{n}=E_{n}(\wp)-E_{n}(\wp_{1})$. Our  goal is so to study the dependence of the cycle efficiency on the modulation of the control parameter, i.e., the electric field amplitude $\wp$.

\begin{center}
\begin{figure}[H]
\centering
\includegraphics*[width=0.55\textwidth]{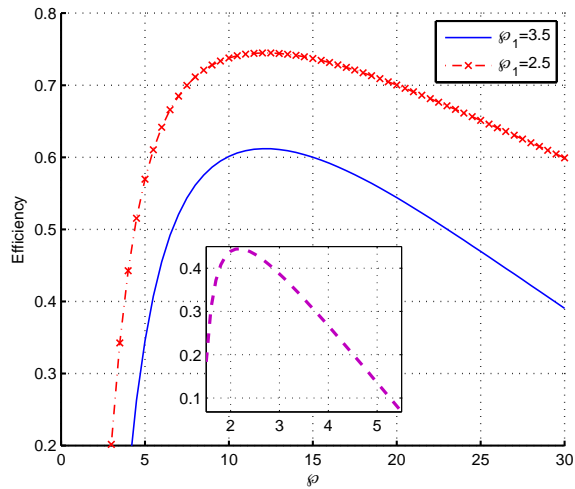}
\caption{(Color online) The efficiency of the quantum Otto cycle as a function of the modulation of the electric field amplitude,
for the following values of the parameters $B=1,~~T_H=30,~~T_L=10$. The inset  corresponds to the quantum adiabatic case.
 As  evident  from
the figure in the case of a thermodynamic adiabatic quantum Otto cycle,
depending on the values of the electric field,  the maximal efficiency reaches the value $\eta = 0.75$ which is higher than  the
maximal efficiency of the Carnot cycle $\eta_{C}=1-T_{L}/T_{H}=0.66$. The efficiency of the
quantum adiabatic Otto cycle (inset) is slightly below the efficiency of the thermodynamic adiabatic Otto cycle.}
\label{Fig.12}
\end{figure}
\end{center}

To this end we considered two slightly different quantum Otto cycles.
 As shown in Fig.\ref{system}, the first cycle consists of  four strokes \cite{Chotorlishvili}: Step $A\rightarrow B$ and $C\rightarrow D$ are two isochoric processes. During the step $A\rightarrow B$ the system couples to the hot bath at temperature $T_{H}$  and the energy levels are unchanged. After absorbing energy from the hot bath, the
system reaches  a thermodynamic equilibrium state, which can be described by the level populations $P_{n}^{A}\big(E_{n}(\wp),T_{H}\big)$. Step
$C\rightarrow D$ is the reverse of the step $A\rightarrow B$.  Namely,  the system is brought to couple to a sink at the temperature
$T_{L}$. After  energy exchange with the heat bath a thermodynamic equilibrium state is established, which can be described by the level populations $P_{n}^{D}\big(E_{n}(\wp_{1}),T_{L}\big)$. Steps $B\rightarrow C$ and $D\rightarrow A$ are  quantum adiabatic processes in which the level populations are unchanged, i.e. $P_{n}^{A}=P_{n}^{D},~~P_{n}^{B}=P_{n}^{C}$. During the process $B\rightarrow C$  amplitude of the electric
field is changed $\Delta E_{n}=E_{n}(\wp)-E_{n}(\wp_{1})$. The working substance performs a positive work.
Therefore, the heat absorbed by the working substance and the heat released read
\cite{Chotorlishvili} $Q_{in}=\displaystyle\sum_{n=1}^{16}E_{n}(\wp)\big(P_{n}^{B}-P_{n}^{A}\big),~~
Q_{out}=\displaystyle\sum_{n=1}^{16}E_{n}(\wp_{1})\big(P_{n}^{B}-P_{n}^{A}\big)$. \\
In the second scenario for the cycle, the quantum adiabatic strokes
of the cycle are replaced by thermodynamic adiabatic strokes. The heat absorbed by the working substance $Q_{in}$
and the heat released in the quantum isochoric cooling process $Q_{out}$ in the case  of the thermodynamic adiabatic cycle
are defined in the following form:
\begin{eqnarray}\label{eq.13}
&Q_{in}=\displaystyle\sum_{n=1}^{16}E_{n}(\wp)\bigg(Z^{-1}(T_{H},\wp)
\exp\bigg[-\frac{E_{n}(\wp)}{T_{H}}\bigg]&\nonumber
\\
&~~~~~~~~~~~~~~~~~~~~~~~~~~-Z^{-1}(T_{L},\wp)\exp\bigg[-\frac{E_{n}(\wp)}{T_{L}}\bigg]\bigg),&
\\
&Q_{out}=\displaystyle\sum_{n=1}^{16}E_{n}(\wp_{1})\bigg(Z^{-1}(T_{H},\wp_{1})
\exp\bigg[-\frac{E_{n}(\wp_{1})}{T_{H}}\bigg]&\nonumber
\\
&~~~~~~~~~~~~~~~~~~~~~~~~~-Z^{-1}(T_{L},\wp_{1})\exp\bigg[-\frac{E_{n}(\wp_{1})}{T_{L}}\bigg]\bigg),&\nonumber
\\ \nonumber
\end{eqnarray}
while the expression for the cycle efficiency reads
\begin{eqnarray}\label{eq.14}
\eta=\frac{Q_{in}-Q_{out}}{Q_{in}}.
\end{eqnarray}
The dependence of the quantum Otto cycle efficiency on the modulation of the electric field amplitude is presented in Fig. \ref{Fig.12}.


\begin{center}
\begin{figure}[H]
\centering
\includegraphics*[width=0.55\textwidth]{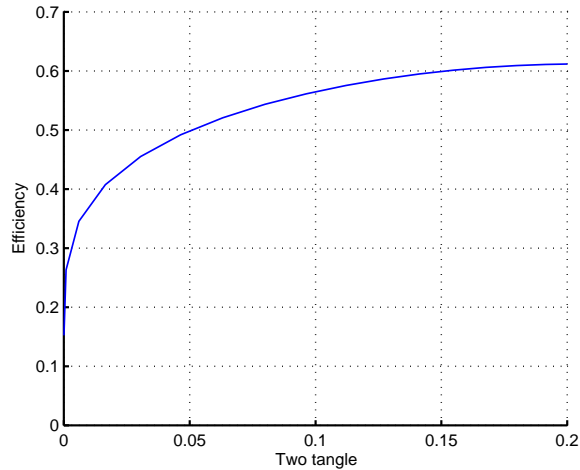}
\caption{(Color online) The efficiency of the quantum Otto cycle as a function of the pair entanglement,
for the following parameters: $\wp_1=3.5,~~B=1$.  We find that a  strong entanglement in the system is related to an enhanced cycle efficiency.}
\label{Fig.13}
\end{figure}
\end{center}

For both types of the cycles (quantum adiabatic and thermodynamic adiabatic) we observed qualitatively similar dependencies  on the
electric field. In both cases the maximal efficiency is reached for certain optimal values of the modulation of the electric field amplitude.
However, the  maximal efficiency obtained for thermodynamic adiabatic cycle is higher compared to the efficiency corresponding to the quantum adiabatic case. Let us concentrate on the thermodynamic adiabatic cycle. As one can see reasonably high efficiency of around $75\%$ is achievable already for $\wp/\wp_{1}\approx5$. We also observe a saturation of the cycle efficiency with a further increase of the electric field amplitude.
Depending on the amplitude of electric field, the efficiency of the quantum Otto engine might  be higher or lower as compared to the maximal efficiency of the Carnot cycle $\eta_{C}=1-T_{L}/T_{H}$. The reason why the efficiency of the quantum cycle exceeds the maximal efficiency of the Carnot cycle is of an entirely quantum origin, as it can be traced back to the entanglement of the working substance \cite{Dillenschneider}. To illustrate this, we plotted the efficiency of the cycle as a function of the entanglement.  Fig. \ref{Fig.13} evidences that an
 increase of the entanglement in the system result in an enhanced cycle efficiency.

\section{Scaling of the cycle efficiency with the size of working substance}

\begin{center}
\begin{figure}[H]
\centering
\includegraphics*[width=0.55\textwidth]{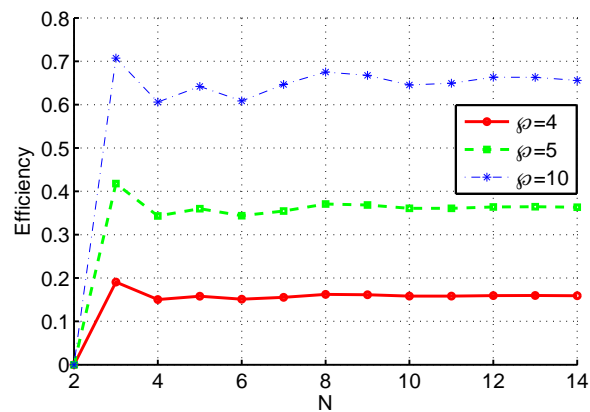}
\caption{(Color online) The efficiency for $\wp_{1}=3.5$ and $\wp=4,~\wp=5,~\wp=10$ is plotted as a function of system size $N$.
 We find that the efficiency jumps abruptly for $N=3$. For $N>4$  no significant
variation in the efficiency is observed when increasing $N$. This is we observed  for all  cases. }
\label{Fig.10N}
\end{figure}
\end{center}

In order to investigate the scaling of the Otto cycle efficiency with the size of the working substance for different values of the electric field we plot the dependence of the cycle efficiency on the length of the spin frustrated MF chain $N$, as shown in Fig.\ref{Fig.10N}.

The case with a  smaller size of the working substance shows a different behavior in comparison to a large system size.
 An abrupt increase in the efficiency is found for $N=3$ . For $N>4$ almost the efficiency hardly change
 with a larger system size. For $N>4$ and a very high electric field $\wp=10$,  we encounter the efficiency corresponding to the saturation value.
\section{Four spins semi-classical limit}
To conclude our analytical considerations we inspect the semi-classical limit utilizing the canonical thermodynamic perturbation  theory \cite{Landau}.
We proceed by writing for the Hamiltonian (\ref{eq.2})
\begin{eqnarray}\label{eq.15}
&\hat{H}=\hat{H_0}+\hat{V},&\nonumber
\\
&\hat{H_0}=-\displaystyle\sum_{i=1}^{N}\vec{\sigma}_i.\vec{\sigma}_{i+1}
+\displaystyle\sum_{i=1}^{N}\vec{\sigma}_i.\vec{\sigma}_{i+2}-B\displaystyle\sum_{i=1}^{N}\sigma_{i}^{z},&
\\
&\hat{V}=\wp\sum_{i}(\vec{\sigma}_{i}\times
\vec{\sigma}_{i+1})_{z}.&\nonumber
\end{eqnarray}
Here again $J_{1}=-J_{2}=J$ ($B\rightarrow\gamma_{e}\hbar B/J,~~\wp\rightarrow g_{ME}|\overrightarrow{\wp}|/J$).

$\wp$ is assumed to be the small parameter. The  eigenvalues and the eigenfunction of  $\hat{H_0}$ are denoted by
$E^{0}_{n},$ $\vert\phi_{n}\rangle$ . We will utilize the normalization condition $\displaystyle\sum_{n}^{}\exp\bigg[\frac{F-E_{n}}{T}\bigg]=1$ where $F=-T\ln\bigg(\displaystyle\sum_{n}^{}\exp\bigg[-\frac{E_n}{T}\bigg]\bigg)$ is the free energy and the left side of the normalization condition is a function of temperature. Taking  the derivative of the normalization condition and after straightforward  algebraic manipulations  we obtain
\begin{eqnarray}\label{eq.16}
&\Delta F=\bigg(\frac{\partial F}{\partial T}\bigg)_{\wp}\delta T+\bigg<\frac{\partial \hat{H}}{\partial \wp}\bigg>\delta \wp
~~~~~~~~S=-\bigg(\frac{\partial F}{\partial T}\bigg)_{\wp}.&
\end{eqnarray}

On the other hand, using the canonical thermodynamic perturbation theory \cite{Landau} in the semi-classical,  high temperature limit $E^{0}_{n}-E^{0}_{m}<T$ we deduce

\begin{eqnarray}\label{eq.17}
&F(T,\wp)=F_{0}(T,0)+\big<V(\wp)\big>&\nonumber
\\
&~~~~~~~~~~~~~~-\frac{1}{2T}\{\displaystyle\sum_{m\neq n}\big<|V_{nm}(\wp)|^2\big>+\big<(V(\wp)-\big<V(\wp)\big>)^{2}\big>\}.&
\end{eqnarray}

\begin{center}
\begin{figure}[H]
\centering
\includegraphics*[width=0.55\textwidth]{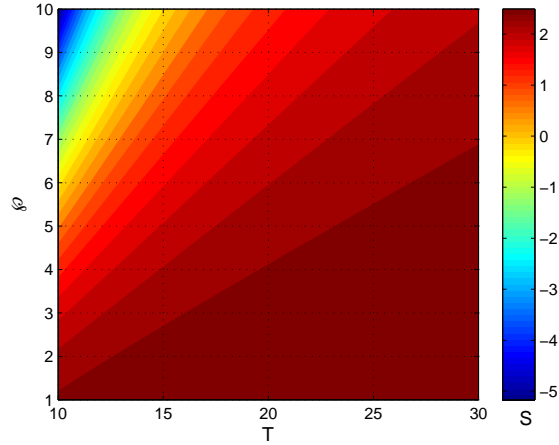}
\caption{(Color online) Contour plot of entropy as a function of the modulation of the electric field amplitude and the temperature. }
\label{Fig.15}
\end{figure}
\end{center}

Here $\big<V_{nn}(\wp)\big>=\displaystyle\sum_{n}P_{n}V_{nn}(\wp)$ is the mean value of the matrix element of the perturbation $\hat{V}$ evaluated in the
basis of the unperturbed Hamiltonian $\hat{H}_{0}$ and $\big<|V_{nm}(\wp)|^2\big>=\displaystyle\sum_{n}P_{n}|V_{nm}(\wp)|^2 $ , $\big<(V(\wp)-\big<V(\wp)\big>)^{2}\big>=\displaystyle\sum_{n}P_{n}{\big(V_{nn}-\sum_{k}P_{k}V_{kk}\big)}^2$. The level populations are described
in terms of the Gibbs distribution function $P_n$. For four spins according to (\ref{eq.17}) we have
\begin{eqnarray}\label{eq.18}
&F(T,\wp)=-T\ln(\displaystyle\sum_{n=1}^{16}e^{-\frac{E^{0}_{n}}{T}})
-\frac{16\wp^{2}(e^{-\frac{E^{0}_{2}}{T}}
+e^{-\frac{E^{0}_{12}}{T}}+e^{-\frac{E^{0}_{6}}{T}}+e^{-\frac{E^{0}_{7}}{T}})}{T\displaystyle\sum_{n=1}^{16}e^{-\frac{E^{0}_{n}}{T}}}.&
\end{eqnarray}
Where $E^{0}_{2}=4J_{2}-2B$, $E^{0}_{12}=4J_{2}+2B$, $E^{0}_{6}=4J_{1}-4J_{2}$ and $E^{0}_{7}=12J_{2}$. For other energy values of the unperturbed Hamiltonian
$\hat{H}_{0}$ we refer to the  supplementary materials.
Combining (\ref{eq.16}), (\ref{eq.17}) one can infer that the change in the entropy during the heating and cooling processes depends not only on the initial and the final temperature but also on the electric field. In detail one finds

\begin{eqnarray}\label{eq.19}
&S(T,\wp)=\ln\bigg(\displaystyle\sum_{n=1}^{16}e^{-\frac{E^{0}_{n}}{T}}\bigg)
+\frac{\displaystyle\sum_{n=1}^{16}E^{0}_{n}e^{-\frac{E^{0}_{n}}{T}}}{T\displaystyle\sum_{n=1}^{16}e^{-\frac{E^{0}_{n}}{T}}}-&\nonumber
\\
&~~~~~~~~~~~~~~16\wp^{2}\bigg(\frac{e^{-\frac{E^{0}_{2}}{T}}
+e^{-\frac{E^{0}_{12}}{T}}+e^{-\frac{E^{0}_{6}}{T}}+e^{-\frac{E^{0}_{7}}{T}}}{T^{2}\displaystyle\sum_{n=1}^{16}e^{-\frac{E^{0}_{n}}{T}}}-&\nonumber
\\
&~~~~~~~~~~~~~~\frac{\big(E^{0}_{2}e^{-\frac{E^{0}_{2}}{T}}
+E^{0}_{12}e^{-\frac{E^{0}_{12}}{T}}+E^{0}_{6}e^{-\frac{E^{0}_{6}}{T}}+E^{0}_{7}e^{-\frac{E^{0}_{7}}{T}}\big)\displaystyle\sum_{n=1}^{16}e^{-\frac{E^{0}_{n}}{T}}}{T^{3}\bigg({\displaystyle\sum_{n=1}^{16}e^{-\frac{E^{0}_{n}}{T}}}\bigg)^2}&\nonumber
\\
&~~~~~~~~~~~~~~-\frac{\big(e^{-\frac{E^{0}_{2}}{T}}
+e^{-\frac{E^{0}_{12}}{T}}+e^{-\frac{E^{0}_{6}}{T}}+e^{-\frac{E^{0}_{7}}{T}}\big)\displaystyle\sum_{n=1}^{16}E^{0}_{n}e^{-\frac{E^{0}_{n}}{T}}}{T^{3}\bigg({\displaystyle\sum_{n=1}^{16}e^{-\frac{E^{0}_{n}}{T}}}\bigg)^2}\bigg).&
\end{eqnarray}
We remark that the entropy is defined by the partial derivative of the free energy with respect to  the temperature at  constant values of the
electric field (\ref{eq.16}). Therefore, the dependence of the entropy on the electric field is parametrical. If the temperature is constant the
entropy is  constant, as well. However, the change in the entropy due to a change of the temperature depends on the values of the electric field.
Eq. (\ref{eq.19}) tells that this dependence is quadratic in the field $S\big(T,\wp\big)-S\big(T,\wp=0\big)\approx \wp^{2}(\cdots)$. The entropy
as a function of the temperature
for different values of the parameters of electric field is plotted at the Fig. \ref{Fig.15}.

The  maximum of the entropy is observed for  small
values of the electric field and in the high temperature limit.
We note that Eq.(\ref{eq.19}) is obtained via a thermodynamic perturbation theory and
negative values of the entropy correspond to  values of the parameters beyond the range where perturbation theory is viable.
Taking into account Eq.(\ref{eq.19}) we can express  the semi-classical efficiency in terms of the electric field as

\begin{eqnarray}\label{eq.20}
\eta_{sc}=1-\frac{\displaystyle\int_{T_L}^{T_H}T\frac{\partial S(T,\wp)}{\partial T}d T}{\displaystyle\int_{T_L}^{T_H}T\frac{\partial S(T,\wp_{1})}{\partial T}d T}.
\end{eqnarray}

\begin{center}
\begin{figure}[H]
\centering
\includegraphics*[width=0.55\textwidth]{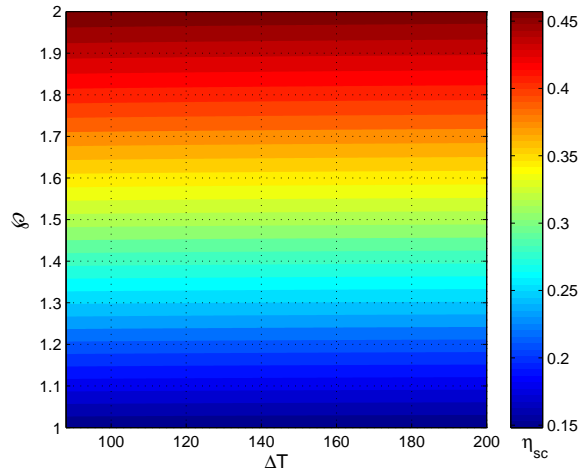}
\caption{(Color online) The semiclassical efficiency $\lambda$ as a function of the electric field $\wp$
and the temperature difference between the hot and the cold baths $\Delta T=T_{H}-T_{L}$ for $T_{L}=100$, $\wp_1=0.5$ plotted using (\ref{eq.20}).
 The semiclassical efficiency
is more sensitive to the values of the electric field  than to the temperature difference $\Delta T$.}
\label{Fig.16}
\end{figure}
\end{center}

The semiclassical efficiency as a function of the electric field $\wp$
and the temperature difference between the hot and the cold baths $\Delta T=T_{H}-T_{L}$ for $T_{L}=100$ are presented  in
Fig. \ref{Fig.16}.

The results exhibit a larger sensitivity of  the semiclassical efficiency to electric field $\wp$ than to   temperature variations $\Delta T$.

\section{Conclusions}

In the present project we studied a quantum Otto engine operating with a working substance consisting of electrically controlled multiferroic
 spin chain.
We have shown that due to the existence of a nonzero spin chirality coupled to an emergent electric polarization in the working substance, the
 efficiency of the cycle is sensitive to the
applied external electric field. We analyzed the dependence of the cycle efficiency on the size of the working substance. In particular,
for a small  working substance consisting of $N=3$ spins, the efficiency reaches considerably  high values, i.e., slightly below  $100\%$.
With  increasing  the size of the working substance, the efficiency of the quantum Otto cycle, shows a saturation plateau. Another interesting
finding is   the robustness of the many-body collective entanglement to temperature variations. Thereby, the many-body entanglement
is quantified in terms of one tangle $\tau_{1}$ and is always larger than the total pair concurrence, as described  by the
two tangle $\tau_{2}$. This indicates that a major amount of the entanglement of the MF working substance is stored in the
long-range, multi-spin correlations (see Fig. \ref{Fig.5}).
In contrast  to the one tangle, the pair correlation is sensitive to the increase in the temperature. In particular, we observe the
existence of a threshold temperature for the two tangle $\tau_{2}$. The stronger the electric field amplitude, the higher is the threshold temperature, as illustrated in Fig. \ref{Fig.2}. The same behavior we encounter for the chirality, as well (cf. Fig. \ref{Fig.6}).
 In particular, with
increasing  the temperature the thermal chirality undergoes  strong changes and for the threshold temperature $T_{C}$
of the two tangle $\tau_{2}$, the thermal chirality reaches its minima. Thus, the thermal chirality
is related rather to the pair correlations in the system, not to the nonlocal entanglement.
We also studied the relations of the magnetic and the electric susceptibilities to the temperature.
According to the definition Eq.(\ref{eq.11}), the maximum of the susceptibilities corresponds to a minimum of the fidelity.
Comparing Fig. \ref{Fig.8} - Fig. \ref{Fig.9N} with Fig. \ref{Fig.2}, we conclude on a direct correlation between the threshold temperature of the pair entanglement
$\tau_{2}$ and the minimum of the fidelities.  The maximum of
the electric and the magnetic susceptibilities are related  to the corresponding threshold temperatures of the pair correlations.
 The maximum of the electric susceptibility is observed almost at the threshold temperatures $T_{c}\approx24,~~T_{c}\approx45$ of the pair entanglement $\tau_{2}$. Indeed, the quantum engine with a MF working substance is much more sensitive to the electric field than to the magnetic field which we
 think is favorable from an experimental point of view.

%
\section{Acknowledgements}
We gratefully acknowledge valuable discussions with
James Anglin. MA, LC, SKM and JB acknowledge gratefully financial
support by the Deutsche Forschungsgemeinschaft (DFG)
through SFB 762, and contract BE 2161/5-1.  TV is
supported by QUEST (Center for Quantum Engineering and
Space-Time Research) and DFG Research Training Group
(Graduiertenkolleg) 1729. WH acknowledges support by Deutsche
Forschungsgemeinschaft through SFB-TR 88.

\newpage
{\fontsize{14}{10}\selectfont \textbf{Supplementary Materials to: Quantum Otto heat engine based on a multiferroic chain working substance}}
\vspace{2cm}

{\textbf{A. Eigenfunctions and eigenvalues}}
\vspace{0.5cm}

Eigenfunctions and corresponding eigenvalues of the Hamiltonian (2) for  four spins are,

\begin{eqnarray}\label{eq.A1}
&&\vert\psi_{1}\rangle=\vert0000\rangle,\nonumber\\
&&\vert\psi_{2}\rangle=\frac{-i}{2}\vert1000\rangle+\frac{-1}{2}\vert0100\rangle
+\frac{i}{2}\vert0010\rangle+\frac{1}{2}\vert0001\rangle,\nonumber\\
&&\vert\psi_{3}\rangle=\frac{i}{2}\vert1000\rangle+\frac{-1}{2}\vert0100\rangle
+\frac{-i}{2}\vert0010\rangle+\frac{1}{2}\vert0001\rangle,\nonumber\\
&&\vert\psi_{4}\rangle=\frac{1}{2}\vert1000\rangle+\frac{-1}{2}\vert0100\rangle
+\frac{1}{2}\vert0010\rangle+\frac{-1}{2}\vert0001\rangle,\nonumber\\
&&\vert\psi_5\rangle=\frac{1}{2}\vert1000\rangle+\frac{1}{2}\vert0100\rangle
+\frac{1}{2}\vert0010\rangle+\frac{1}{2}\vert0001\rangle,\nonumber\\
&&\vert\psi_6\rangle=\alpha\big(\vert1100\rangle-i\eta\vert1010\rangle
-\vert1001\rangle-\vert0110\rangle
+i\eta\vert0101\rangle+\vert0011\rangle\big),\nonumber\\
&&\vert\psi_7\rangle=\gamma\big(\vert1100\rangle-i\lambda\vert1010\rangle
-\vert1001\rangle-\vert0110\rangle
+i\lambda\vert0101\rangle+\vert0011\rangle\big),\nonumber\\
&&\vert\psi_8\rangle=\frac{1}{\sqrt{6}}\big(\vert1100\rangle+\vert1010\rangle
+\vert1001\rangle+\vert0110\rangle
+\vert0101\rangle+\vert0011\rangle\big),\nonumber\\
&&\vert\psi_9\rangle=\frac{1}{\sqrt{12}}\big(\vert1100\rangle-2\vert1010\rangle
+\vert1001\rangle+\vert0110\rangle
-2\vert0101\rangle+\vert0011\rangle\big),\nonumber\\
&&\vert\psi_{10}\rangle=\frac{-1}{\sqrt{2}}\vert1100\rangle
+\frac{1}{\sqrt{2}}\vert0011\rangle,~~~~~~~~~~~~~~~~~~~~~~~~~~~~~~~~~~~~~~~~~~~~~~~~(A1)\nonumber\\
&&\vert\psi_{11}\rangle=\frac{-1}{\sqrt{2}}\vert1001\rangle
+\frac{1}{\sqrt{2}}\vert0110\rangle,\nonumber\\
&&\vert\psi_{12}\rangle=\frac{i}{2}\vert1110\rangle+\frac{-1}{2}\vert1101\rangle
+\frac{-i}{2}\vert1011\rangle+\frac{1}{2}\vert0111\rangle,\nonumber\\
&&\vert\psi_{13}\rangle=\frac{-i}{2}\vert1110\rangle+\frac{-1}{2}\vert1101\rangle
+\frac{i}{2}\vert1011\rangle+\frac{1}{2}\vert0111\rangle,\nonumber\\
&&\vert\psi_{14}\rangle=\frac{1}{2}\vert1110\rangle+\frac{1}{2}\vert1101\rangle
+\frac{1}{2}\vert1011\rangle+\frac{1}{2}\vert0111\rangle,\nonumber\\
&&\vert\psi_{15}\rangle=\frac{1}{2}\vert1110\rangle+\frac{-1}{2}\vert1101\rangle
+\frac{1}{2}\vert1011\rangle+\frac{-1}{2}\vert0111\rangle,\nonumber\\
&&\vert\psi_{16}\rangle=\vert1111\rangle,\nonumber\\
&&E_{1}=-4J_1-4J_2-4B,E_{2}=4J_2-2B-4d,\nonumber\\
&&E_{3}=4J_2-2B+4d,E_{4}=4J_1-4J_2-2B,\nonumber\\
&&E_{5}=-4J_1-4J_2-2B,\nonumber\\
&&E_6=2J_1+4J_2+2\sqrt{J_1^2+16J_2^2-8J_1J_2+8d^2},\nonumber\\
&&E_7=2J_1+4J_2-2\sqrt{J_1^2+16J_2^2-8J_1J_2+8d^2},\nonumber\\
&&E_8=-4J_1-4J_2,E_9=8J_1-4J_2,\nonumber\\
&&E_{10}=E_{11}=4J_2,\nonumber\\
&&E_{12}=4J_2+2B+4d,E_{13}=4J_2+2B-4d,\nonumber\\
&&E_{14}=-4J_1-4J_2+2B,E_{15}=4J_1-4J_2+2B,\nonumber\\
&&E_{16}=-4J_1-4J_2+4B.\nonumber
\end{eqnarray}

Thereby the following notations are used

\begin{eqnarray}\label{eq.A2}
&&\alpha=\frac{1}{\sqrt{4+2\mu^2}},\nonumber\\
&&\mu=\frac{4J_2-J_1
-\sqrt{J_1^2+16J_2^2-8J_1J_2+8d^2}}{2d},~~~~~~~~~~~~~~~~~~~~~~~~~~~~(A2)\nonumber\\
&&\gamma=\frac{1}{\sqrt{4+2\lambda^2}},\nonumber\\
&&\lambda=\frac{4J_2-J_1
+\sqrt{J_1^2+16J_2^2-8J_1J_2+8d^2}}{2d}.\nonumber
\end{eqnarray}
Note, since the magnetic field couples to a conserved quantity,  the eigenfunctions do not depend on the magnetic field amplitude $B$.

The expressions for the parameters entering in the (2)-(7) read,
\begin{eqnarray}\label{eq.A3}
&a_1=e^{-\beta E_1}+\frac{1}{2}(e^{-\beta E_2}+e^{-\beta E_3}+e^{-\beta E_4}+e^{-\beta E_5})&\nonumber\\
&~~~~~~~~+\alpha^2e^{-\beta E_6}+\gamma^2e^{-\beta E_7}+\frac{1}{6}e^{-\beta E_8}
+\frac{1}{12}e^{-\beta E_9}+\frac{1}{2}e^{-\beta E_{10}},&\nonumber
\end{eqnarray}

\begin{eqnarray}\label{eq.A3}
&b_1=\frac{1}{4}(e^{-\beta E_2}+e^{-\beta E_3}+e^{-\beta E_4}+e^{-\beta E_5})&\nonumber\\
&~~~~~~~~+\alpha^2(1+\mu^2)e^{-\beta E_6}+\gamma^2(1+\lambda^2)e^{-\beta E_7}&\nonumber\\
&~~~~~~~~+\frac{1}{3}e^{-\beta E_8}+\frac{5}{12}e^{-\beta E_9}+\frac{1}{2}e^{-\beta E_{11}}&\nonumber\\
&~~~~~~~~+\frac{1}{4}(e^{-\beta E_{12}}+e^{-\beta E_{13}}+e^{-\beta E_{14}}+e^{-\beta E_{15}}),&\nonumber
\end{eqnarray}

\begin{eqnarray}\label{eq.A3}
&c_1=\frac{i}{4}(e^{-\beta E_2}-e^{-\beta E_3})-\frac{1}{4}(e^{-\beta E_4}-e^{-\beta E_5})&\nonumber\\
&~~~~~~~~+2i\alpha^2\mu e^{-\beta E_6}+2i\gamma^2\lambda e^{-\beta E_7}&\nonumber\\
&~~~~~~~~+\frac{1}{3}(e^{-\beta E_8}-e^{-\beta E_9})-\frac{i}{4}(e^{-\beta E_{12}}-e^{-\beta E_{13}})&\nonumber\\
&~~~~~~~~+\frac{1}{4}(e^{-\beta E_{14}}-e^{-\beta E_{15}}),~~~~~~~~~~~~~~~~~~~~~~~~~~~~~~~~~~~~~~~~~~~~~~~(A3)\nonumber
\end{eqnarray}

\begin{eqnarray}\label{eq.A3}
&d_1=\alpha^2e^{-\beta E_6}+\gamma^2e^{-\beta E_7}+\frac{1}{6}e^{-\beta E_8}
+\frac{1}{12}e^{-\beta E_9}+\frac{1}{2}e^{-\beta E_{10}}&\nonumber\\
&~~~~~~~~+\frac{1}{2}(e^{-\beta E_{12}}+e^{-\beta E_{13}}+e^{-\beta E_{14}}+e^{-\beta E_{15}})
+e^{-\beta E_{16}}.&\nonumber
\end{eqnarray}

and

\begin{eqnarray}\label{eq.A4}
&a_2=e^{-\beta E_1}+\frac{1}{2}(e^{-\beta E_2}+e^{-\beta E_3}+e^{-\beta E_4}+e^{-\beta E_5})&\nonumber\\
&~~~~~~~~+\alpha^2\mu^2e^{-\beta E_6}+\gamma^2\lambda^2e^{-\beta E_7}
+\frac{1}{6}e^{-\beta E_8}+\frac{1}{3}e^{-\beta E_9},&\nonumber
\end{eqnarray}

\begin{eqnarray}\label{eq.A4}
&b_2=\alpha^2\mu^2e^{-\beta E_6}+\gamma^2\lambda^2e^{-\beta E_7}+\frac{1}{6}e^{-\beta E_8}
+\frac{1}{3}e^{-\beta E_9}&\nonumber\\
&~~~~~~~~+\frac{1}{2}(e^{-\beta E_{12}}+e^{-\beta E_{13}}+e^{-\beta E_{14}}+e^{-\beta E_{15}})
+e^{-\beta E_{16}},&\nonumber
\end{eqnarray}

\begin{eqnarray}\label{eq.A4}
&c_2=\frac{1}{4}(e^{-\beta E_2}+e^{-\beta E_3}+e^{-\beta E_4}+e^{-\beta E_5})&\nonumber\\
&~~~~~~~~+2\alpha^2e^{-\beta E_6}+2\gamma^2e^{-\beta E_7}+\frac{1}{3}e^{-\beta E_8}&\nonumber\\
&~~~~~~~~+\frac{1}{6}e^{-\beta E_9}+\frac{1}{2}e^{-\beta E_{10}}+\frac{1}{2}e^{-\beta E_{11}}~~~~~~~~~~~~~~~~~~~~~~~~~~~~~~~~~~(A4)&\nonumber\\
&~~~~~~~~+\frac{1}{4}(e^{-\beta E_{12}}+e^{-\beta E_{13}}+e^{-\beta E_{14}}+e^{-\beta E_{15}}),&\nonumber
\end{eqnarray}

\begin{eqnarray}\label{eq.A4}
&d_2=\frac{-1}{4}(e^{-\beta E_2}+e^{-\beta E_3})+\frac{1}{4}(e^{-\beta E_4}+e^{-\beta E_5})&\nonumber\\
&~~~~~~~~-2\alpha^2e^{-\beta E_6}-2\gamma^2e^{-\beta E_7}
+\frac{1}{3}(e^{-\beta E_8}+e^{-\beta E_9})&\nonumber\\
&~~~~~~~~-\frac{1}{4}(e^{-\beta E_{12}}+e^{-\beta E_{13}})-\frac{1}{4}(e^{-\beta E_{14}}-e^{-\beta E_{15}}).&\nonumber
\end{eqnarray}

The parameter $Q$ entering in the expression of the one tangle reads:

\begin{eqnarray}\label{eq.A5}
&Q=\{e^{-\beta E_1}+\frac{3}{4}(e^{-\beta E_2}+e^{-\beta E_3}+e^{-\beta E_4}+e^{-\beta E_5})&\nonumber\\
&~~~~~~+\alpha^2(2+\mu^2)e^{-\beta E_6}+\gamma^2(2+\lambda^2)e^{-\beta E_7}&\nonumber\\
&~~~~~~+\frac{1}{2}(e^{-\beta E_8}+e^{-\beta E_9}+e^{-\beta E_{10}}+e^{-\beta E_{11}})~~~~~~~~~~~~~~~~~~~~~~~~~~(A5)&\nonumber\\
&~~~~~~+\frac{1}{4}(e^{-\beta E_{12}}+e^{-\beta E_{13}}+e^{-\beta E_{14}}+e^{-\beta E_{15}})\}\times&\nonumber\\
&~~~~~~\{\frac{1}{4}(e^{-\beta E_2}+e^{-\beta E_3}+e^{-\beta E_4}+e^{-\beta E_5})+\alpha^2(2+\mu^2)e^{-\beta E_6}&\nonumber\\
&~~~~~~+\gamma^2(2+\lambda^2)e^{-\beta E_7}+\frac{1}{2}(e^{-\beta E_8}+e^{-\beta E_9}+e^{-\beta E_{10}}+e^{-\beta E_{11}})&\nonumber\\
&~~~~~~+\frac{3}{4}(e^{-\beta E_{12}}+e^{-\beta E_{13}}+e^{-\beta E_{14}}+e^{-\beta E_{15}})+e^{-\beta E_{16}}\}.&\nonumber
\end{eqnarray}

The susceptibilities of the system with respect to the external magnetic and electric fields are

\begin{eqnarray}\label{eq.A6}
&\chi(B)=-\frac{\partial^2 F}{\partial B^2}=&\nonumber\\
&~~~~~~~~\frac{\displaystyle\sum_{n,m=1}^{16}[\beta(\frac{dE_n}{dB})^2
-\beta(\frac{dE_n}{dB})(\frac{dE_m}{dB})-\frac{d^2E_n}{dB^2}]\exp[-\beta(E_n+E_m)]}
{(\displaystyle\sum_{n=1}^{16}\exp[-\beta E_n])^2}&\nonumber\\
&=\frac{4\beta}{Z^2}[e^{-\beta E_1}(e^{-\beta E_2}+e^{-\beta E_3}+e^{-\beta E_4}+e^{-\beta E_5}+4e^{-\beta E_6}&\nonumber\\
&~~~~~~+4e^{-\beta E_7}+4e^{-\beta E_8}+4e^{-\beta E_9}+4e^{-\beta E_{10}}+4e^{-\beta E_{11}}&\nonumber\\
&~~~~~~+9e^{-\beta E_{12}}+9e^{-\beta E_{13}}+9e^{-\beta E_{14}}+9e^{-\beta E_{15}})&\nonumber\\
&~~~~~~+e^{-\beta E_{16}}(9e^{-\beta E_2}+9e^{-\beta E_3}+9e^{-\beta E_4}+9e^{-\beta E_5}&\nonumber\\
&~~~~~~+4e^{-\beta E_6}+4e^{-\beta E_7}+4e^{-\beta E_8}+4e^{-\beta E_9}+4e^{-\beta E_{10}}&\nonumber\\
&~~~~~~+4e^{-\beta E_{11}}+e^{-\beta E_{12}}+e^{-\beta E_{13}}+e^{-\beta E_{14}}+e^{-\beta E_{15}})&\nonumber\\
&~~~~~~+(e^{-\beta E_2}+e^{-\beta E_3}+e^{-\beta E_4}+e^{-\beta E_5}+e^{-\beta E_{12}}&\nonumber\\
&~~~~~~+e^{-\beta E_{13}}+e^{-\beta E_{14}}+e^{-\beta E_{15}})(e^{-\beta E_6}&\nonumber\\
&~~~~~~e^{-\beta E_7}+e^{-\beta E_8}+e^{-\beta E_9}+e^{-\beta E_{10}}+e^{-\beta E_{11}})&\nonumber\\
&~~~~~~+4(e^{-\beta E_2}+e^{-\beta E_3}+e^{-\beta E_4}+e^{-\beta E_5})&\nonumber\\
&~~~~~~(e^{-\beta E_{12}}+e^{-\beta E_{13}}+e^{-\beta E_{14}}+e^{-\beta E_{15}})].~~~~~~~~~~~~~~~~~~~~~~~~~~~(A6)&\nonumber\\&\nonumber
\end{eqnarray}

\begin{eqnarray}\label{eq.A7}
&\chi(\wp)=-\frac{\partial^2 F}{\partial \wp^2}=&\nonumber\\
&~~~~\frac{\displaystyle\sum_{n,m=1}^{16}[\beta(\frac{dE_n}{d\wp})^2
-\beta(\frac{dE_n}{d\wp})(\frac{dE_m}{d\wp})-\frac{d^2E_n}{d\wp^2}]\exp[-\beta(E_n+E_m)]}
{(\displaystyle\sum_{n=1}^{16}\exp[-\beta E_n])^2}&\nonumber\\
&~~~~~=\frac{16\beta}{Z^2}[(e^{-\beta E_2}+e^{-\beta E_{13}})(e^{-\beta E_1}+2e^{-\beta E_3}+e^{-\beta E_4}+e^{-\beta E_5}&\nonumber\\
&~~~~~~+(1+\frac{4\wp}{\sqrt{(5J)^2+8\wp^2}})e^{-\beta E_6}+(1-\frac{4\wp}{\sqrt{(5J)^2+8\wp^2}})e^{-\beta E_7}&\nonumber\\
&~~~~~~+e^{-\beta E_8}+e^{-\beta E_9}+e^{-\beta E_{10}}+e^{-\beta E_{11}}+2e^{-\beta E_{12}}&\nonumber\\
&~~~~~~+e^{-\beta E_{14}}+e^{-\beta E_{15}}+e^{-\beta E_{16}})+(e^{-\beta E_{3}}+e^{-\beta E_{12}})&\nonumber\\
&~~~~~~(e^{-\beta E_1}+2e^{-\beta E_2}+e^{-\beta E_4}+e^{-\beta E_5}&\nonumber\\
&~~~~~~+(1-\frac{4\wp}{\sqrt{(5J)^2+8\wp^2}})e^{-\beta E_6}+(1+\frac{4\wp}{\sqrt{(5J)^2+8\wp^2}})e^{-\beta E_7}&\nonumber\\
&~~~~~~+e^{-\beta E_8}+e^{-\beta E_9}+e^{-\beta E_{10}}+e^{-\beta E_{11}}+2e^{-\beta E_{13}}&\nonumber\\
&~~~~~~+e^{-\beta E_{14}}+e^{-\beta E_{15}}+e^{-\beta E_{16}})&\nonumber\\
&~~~~~~+(\frac{16\wp^2}{(5J)^2+8\wp^2}-\frac{(5J)^2}{\beta((5J)^2+8\wp^2)^{\frac{3}{2}}})e^{-\beta E_6}(e^{-\beta E_1}+e^{-\beta E_4}&\nonumber\\
&~~~~~~+e^{-\beta E_5}+e^{-\beta E_8}+e^{-\beta E_9}+e^{-\beta E_{10}}+e^{-\beta E_{11}}&\nonumber\\
&~~~~~~+e^{-\beta E_{14}}+e^{-\beta E_{15}}+e^{-\beta E_{16}})+(\frac{16\wp^2}{(5J)^2+8\wp^2}&\nonumber\\
&~~~~~~+\frac{4\wp}{\sqrt{(5J)^2+8\wp^2}}-\frac{(5J)^2}{\beta((5J)^2+8\wp^2)^{\frac{3}{2}}})e^{-\beta E_6}(e^{-\beta E_2}+e^{-\beta E_{13}})&\nonumber\\
&~~~~~~+(\frac{16\wp^2}{(5J)^2+8\wp^2}-\frac{4\wp}{\sqrt{(5J)^2+8\wp^2}}-\frac{(5J)^2}{\beta((5J)^2+8\wp^2)^{\frac{3}{2}}})e^{-\beta E_6}(&\nonumber\\
&~~~~~~e^{-\beta E_3}+e^{-\beta E_{12}})-\frac{(5J)^2}{\beta((5J)^2+8\wp^2)^{\frac{3}{2}}})e^{-2\beta E_6}+&\nonumber\\
&~~~~~~(\frac{32\wp^2}{(5J)^2+8\wp^2}-\frac{(5J)^2}{\beta((5J)^2+8\wp^2)^{\frac{3}{2}}})e^{-\beta E_6}e^{-\beta E_7}+&\nonumber\\
&~~~~~~(\frac{16\wp^2}{(5J)^2+8\wp^2}+\frac{(5J)^2}{\beta((5J)^2+8\wp^2)^{\frac{3}{2}}})e^{-\beta E_7}(
e^{-\beta E_1}+e^{-\beta E_4}+e^{-\beta E_5}+&\nonumber\\
&~~~~~~e^{-\beta E_8}+e^{-\beta E_9}+e^{-\beta E_{10}}+e^{-\beta E_{11}}+e^{-\beta E_{14}}+e^{-\beta E_{15}}+e^{-\beta E_{16}})&\nonumber\\
&~~~~~~+(\frac{16\wp^2}{(5J)^2+8\wp^2}-\frac{4\wp}{\sqrt{(5J)^2+8\wp^2}}+\frac{(5J)^2}{\beta((5J)^2+8\wp^2)^{\frac{3}{2}}})e^{-\beta E_7}(&\nonumber\\
&~~~~~~e^{-\beta E_{2}}+e^{-\beta E_{13}})+(\frac{16\wp^2}{(5J)^2+8\wp^2}+\frac{4\wp}{\sqrt{(5J)^2+8\wp^2}}&\nonumber\\
&~~~~~~+\frac{(5J)^2}{\beta((5J)^2+8\wp^2)^{\frac{3}{2}}})e^{-\beta E_7}(e^{-\beta E_{3}}+e^{-\beta E_{12}})+&\nonumber\\
&~~~~~~(\frac{32\wp^2}{(5J)^2+8\wp^2}+\frac{(5J)^2}{\beta((5J)^2+8\wp^2)^{\frac{3}{2}}})e^{-\beta E_7}e^{-\beta E_6}+&\nonumber\\
&~~~~~~\frac{(5J)^2}{\beta((5J)^2+8\wp^2)^{\frac{3}{2}}})e^{-2\beta E_7}].~~~~~~~~~~~~~~~~~~~~~~~~~~~~~~~~~~~~~~~~~(A7)&\nonumber
\end{eqnarray}

Eigenvalues (and eigenfunctions) of the unperturbed Hamiltonian $\hat H_0$ (15) in the case of four spins for zero electric field can be obtained from Eqs. (A1) by putting $d=0$.

\end{document}